\documentclass[11pt]{article}
\usepackage{amsmath}
\usepackage{amsfonts}
\usepackage{mathrsfs}
\usepackage{pstricks}
\usepackage{color}
\usepackage{graphicx}
\usepackage{slashed}
\usepackage{amssymb}

\newcommand{\be}{\begin{equation}}
\newcommand{\ee}{\end{equation}}
\newcommand{\ba}{\begin{array}{c}}
\newcommand{\ea}{\end{array}}
\newcommand{\bqa}{\begin{eqnarray}}
\newcommand{\eqa}{\end{eqnarray}}
\newcommand{\bm}[1]{\mbox{\boldmath{$#1$}}}
\makeatletter
    
    \newcommand{\Rmnum}[1]{\expandafter\@slowromancap\romannumeral #1@}
    \makeatother

\begin{document}

\begin{center}
{\Large\bf $N\bar N$ Scattering at NLO Order in An Effective Theory}
\end{center}
\vskip 10mm
\begin{center}
G.Y. Chen$^1$ and J.P. Ma$^{2,3}$    \\
{\small {\it $^1$ Department of Physics, Peking University,
Beijing 100871, China }} \\
{\small {\it $^2$ Institute of Theoretical Physics, Academia Sinica,
Beijing 100190, China }} \\
{\small {\it $^3$ Center for High-Energy Physics, Peking University, Beijing 100871, China }} \\
\end{center}
\begin{abstract}
We have proposed to use an effective theory to describe interactions of an $N\bar N$-system.
The effective theory can be constructed in analogy to the existing effective theory
for an $NN$-system.
In this work we study the next-to-leading order correction
to $N\bar N$ scattering near the threshold in the effective theory.
We find that the experimental data can be well described with the effective theory.
\end{abstract}

\par\vskip20pt
\par

Interactions between a nucleon and an antinucleon have been studied extensively
with potential models(see \cite{Nimegen,Paris,Julich} and references therein).
In general
the potential of these models includes various effects of  meson exchanges and is
supplemented with a phenomenologically determined imaginary part to account for the
$N\bar N$ annihilation. The descriptions of the low energy $N\bar N$
scattering with these potential models are successful.
Recently there is a renewed interest to study $N\bar N$ interactions,
stimulated by experimental observations of the threshold enhancement
of a $p\bar p$ system in the radiative decay of $J/\psi$\cite{BES},
$B$-decays\cite{Bell} and $e^+e^-$-annihilations\cite{Bar}.
It has been shown that the enhancement can be explained with the final state interaction,
where the enhancement is determined by the $N\bar N$ scattering amplitude near the threshold
\cite{Tbes2,Tbes3,Tbes4,HML,Tbell,Tbar1,Tbar2,Tbar3,CDM,CDM2}.
\par
Because the momentum transfer in $N\bar N$ scattering near the threshold
is small, the inner structure of a nucleon or antinucleon can not be seen.
This indicates that one can take the nucleon and antinucleon as point-like
particles and construct an effective theory to describe the $N\bar N$ scattering
near the threshold. Near the threshold the interactions through $\pi$-exchanges
are most important among those of exchanges of other mesons. hence, the effective theory
only contains $N$-, $\bar N$- and $\pi$-fields.
In \cite{CDM2} we have proposed such an effective theory.
We have used the tree-level result of the effective theory and a partial sum of higher
order effects to successfully describe the observed enhancement and the $N\bar N$ scattering near the threshold.
In this work we study the next-to-leading order correction in the effective theory.
\par
An effective theory for an $N\bar N$ system near the threshold can be constructed
in analogy to the effective theory for an $NN$ system.
The effective theory for the $NN$ system
has been proposed in\cite{Wein,Van,KSW} and studied extensively in
\cite{Van,KSW,Fleming}. In constructing such  an effective theory one
makes an power expansion in the momentum near the threshold.
In the effective theory, the U.V. divergences are regularized with the dimensional regularization.
The interactions with $\pi$-mesons are fixed with chiral symmetry.
A power counting to determine the relative importance of different terms in
the effective theory has to be established.
But,
there are distinct difference between the effective theory of $N\bar
N$ interactions and that of $NN$ interactions.
\par
It is well-known that for an $NN$ system the scattering lengthes
are large and there is a shallow bound state in
$^3S_1-{^3}D_1$ channel and a virtual state in $^1S_0$ channel.
To take these facts into account in the effective theory, the power subtraction scheme
has been introduced\cite{KSW}. With this scheme a systematic power counting
of the effective theory can be established.
In the case of $N\bar N$ systems, the scattering
lengthes, according LEAR experiment\cite{Tbes2} and model
results\cite{Tbes3}, are around 1fm. They are much
smaller than those of $NN$ systems. Therefor one can use the minimal
subtraction scheme and hence the simple power counting for the
effective theory of $N\bar N$ systems.
\par
Another difference is that an $N\bar N$ system can be annihilated
into mesons, while a $NN$ system can not be annihilated. The
annihilation of an $N\bar N$ system into virtual or real mesons
results in that the dispersive and absorptive part of the $N\bar N$
scattering amplitudes are relatively of the same importance. In
order to incorporate this fact some coupling constants in the
effective theory of an $N\bar N$ system are complex numbers. One
should keep in mind that the complex coupling constants here do not
mean the violation of time-reversal symmetry.
The effective Lagrangian with complex coupling
constants should be understood as for the purpose to effectively
build the $S$-operator and hence scattering amplitudes.
This can be understood
as the following: One can imagine that the effective theory is
obtained from a perturbative matching of a more fundamental theory.
In the more fundamental theory with the time-reversal symmetry
scattering amplitudes can become complex beyond tree-level because
absorptive parts can exist at one- or more loop level. The imaginary
parts of the coupling constants in the effective theory
are from these absorptive parts in
the matching. If the underlying theory respects to the time-reversal symmetry and
other discrete symmetry like charge-conjugation and parity,
the generated operators through the matching in the effective theory
can not violate these discrete symmetries of the underlying theory.
This fact also tell us that the effective theory should be built
with these operators which are $C$-, $P$- and $T$-even.
\par
At low energy, i.e., for an $N\bar N$ system near its threshold, we
need to consider theory contains contact interactions and
interactions with pions which is consistent with chiral symmetry. At
low energy we can use the nonrelativistic fields to describe the
nucleon $N$. The nonrelativistic fields are given as
\begin{equation}
\psi=\left(
       \begin{array}{c}
         \psi_p \\
         \psi_n \\
       \end{array}
     \right),\ \ \ \ \ \
     \chi=\left(
            \begin{array}{c}
              \chi_p \\
              \chi_n \\
            \end{array}
          \right).
          \end{equation}
The field $\psi(\chi^\dagger)$ annihilates a
nucleon(an antinucleon) and the field $\psi^\dagger(\chi)$
creates a nucleon(an antinucleon). The pion-field is defined with
Pauli matrices $\tau^i$ acting in the isospin space as:
\begin{equation}
\xi(x)=e^{i\frac{M}{2f_\pi}},\ \ \ \ M=\tau^i\pi^i=\left(
                                             \begin{array}{cc}
                                               \pi^3 & \pi^1-i\pi^2 \\
                                               \pi^1+i\pi^2 & -\pi^3 \\
                                             \end{array}
                                           \right),
                                           \end{equation}
with $f_\pi\approx93$ MeV. The field $\pi^i(i=1,2,3)$ is real. Under
$SU(2)_L\times SU(2)_R$ chiral symmetry the fields transform as
\begin{equation}
\xi\rightarrow L\xi U^\dagger=U \xi R^\dagger,\ \
\Sigma\equiv\xi^2\rightarrow L\Sigma R^\dagger,\ \ \psi\rightarrow
U\psi,
\end{equation}
where L, R are global transformations in $SU(2)_L$ and $SU(2)_R$
respectively and $U$ is a pion-field-dependent SU(2) matrix. From $\xi$
we can give out the vector and axial-vector pion currents as
\begin{equation}
V_\mu=\frac{1}{2}(\xi\partial_\mu\xi^\dagger+\xi^\dagger\partial_\mu\xi),\
\ \
A_\mu=\frac{i}{2}(\xi\partial_\mu\xi^\dagger-\xi^\dagger\partial_\mu\xi),
\end{equation}
where the axial current $A_\mu$ and the chiral covariant derivative
$D_\mu=(\partial_\mu+V_\mu)$ transform linearly as
\begin{equation}
A_\mu\rightarrow UA_\mu U^\dagger,\ \ \ D_\mu\rightarrow UD_\mu
U^\dagger.
\end{equation}
\par
With the above fields the effective Lagrangian can be
constructed as:
\begin{eqnarray}
\mathcal{L}_{eff}&=&\mathcal{L}_N+\mathcal{L}_{\pi},\nonumber\\
\mathcal{L}_{N}&=&\psi^\dagger(iD_0+{\bf{D}}^2/2m)\psi+\chi^\dagger(iD_0-{\bf{D}}^2/2m)\chi+
\frac{c_0}{4}\psi^\dagger\chi\chi^\dagger\psi+\frac{c_1}{4}\psi^\dagger\tau^i\chi\chi^\dagger
\tau^i\psi\nonumber\\
&&+\frac{d_0}{4}\psi^\dagger\sigma^i\chi\chi^\dagger\sigma^i\psi+\frac{d_1}{4}\psi^\dagger\tau^i\sigma^j
\chi\chi^\dagger\sigma^j\tau^i\psi+g_A\psi^\dagger\bm{\sigma}\cdot{\bf{A}}\psi+g_A\chi^\dagger\bm{\sigma}\cdot{\bf{A}}\chi+\mathcal{O}(p^2),\nonumber\\
\mathcal{L}_{\pi}&=&\frac{f_\pi^2}{4}\mbox{Tr}\partial_\mu\Sigma^\dagger\partial^\mu\Sigma+\frac{m_\pi^2
f_\pi^2}{4}\mbox{Tr}(\Sigma+\Sigma^\dagger)+\mathcal{O}(p^4),\label{Lagrangian}
\end{eqnarray}
where $g_A\approx1.25$. $c_I$ with the isospin index $I$ is the coupling constant in the
$^1S_0$ channel, while $d_I$ is the coupling constant
in the $^3S_1$ channel. These coupling constants
are in general complex to account for the $N\bar N$ annihilation.
\par
The power counting of the effective theory is discussed in detail in\cite{CDM2}.
It is easy from the effective Lagrangian to derive the tree-level $N\bar N$ scattering amplitude.
The amplitude is an expansion in the three momentum $p$ of $N$ or $\bar N$. The leading order is at order of $p^0$.
the amplitude at this order is determined by those contact terms given explicitly in Eq.(6) and the interaction
from exchange of single $\pi$.
We will consider the momentum region $p\sim m_\pi$. From Eq.(6),
the $NN\pi$- or $\bar N\bar N \pi $ vertex is proportional to $p$. In the single $\pi$-exchange,
the power of $p$ from the two vertices is canceled by the power of the denominator of the $\pi$-propagator.
This is why the amplitude with single $\pi$-exchange is taken at order of $p^0$.
The next-to-leading order is at order of $p$ and
is determined by one-loop diagrams. It should be noted that some of one-loop diagrams will not contribute
at the order. They can contribute at order of $p^2$ or higher.
As we will show that it is easy to find those one-loop diagrams which give the contributions
at order of $p$. At order of $p^2$ the amplitude receives corrections from different sources.
The corrections can come from those one-loop diagrams which contribute at order of $p^2$ and
come from two-loop diagrams. They also come from those contact terms in the effective Lagrangian
at order of $p^2$, which we have not included in Eq.(6). Those contact terms in general contain
derivatives. We will use the dimensional regularization and the minimal subtraction scheme as discussed before.

\par
\begin{figure}[hbt]
\begin{center}
  \includegraphics[width=7cm]{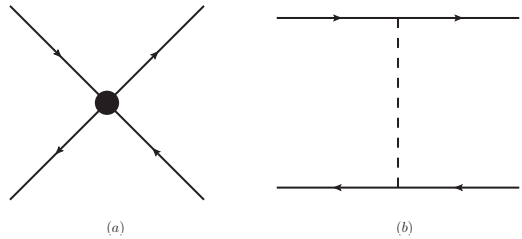}\\
  \caption{The leading order diagrams. Fig.1a is for the contact terms, Fig.1b is from one $\pi$-exchange.}\label{fig1}
  \end{center}
\end{figure}
\par
After giving the effective theory, we now consider the $N\bar N$ scattering near the threshold:
\begin{equation}
N(\vec{p},s_1)+\bar N(-\vec{p},s_2)\rightarrow
N(\vec{k},s_1^\prime)+\bar N(-\vec{k},s_2^\prime),\ \ \
|\vec{p}|=p=\beta\sqrt{m^2+p^2},\label{beta}
\end{equation}
where $\vec{p}$ and $\vec{k}$ are three-momentum. The spins are
denoted as $s^,$s. $\beta$ is the velocity. $m$ is the mass of nucleons.
As discussed before, the
leading order contribution is at $\mathcal{O}(p^0)$. The
contributions at this order can be represented by diagrams in Fig.1.
The leading order amplitude can be expressed as,
\begin{eqnarray}
\mathcal{T}^{(0)}&=&\frac{c_0}{4}\xi^\dagger\eta\eta^\dagger\xi
+\frac{c_1}{4}\xi^\dagger\tau^a\eta\eta^\dagger\tau^a\xi
+\frac{d_0}{4}\xi^\dagger\sigma^i\eta\eta^\dagger\sigma^i\xi+
\frac{d_1}{4}\xi^\dagger\sigma^i\tau^a\eta\eta^\dagger\sigma^i\tau^a\xi\nonumber\\
&&+(\frac{g_A}{2f_\pi})^2\xi^\dagger\vec{\sigma}\cdot(\vec{p}-\vec{k})\tau^a\xi\eta^\dagger\vec{\sigma}
\cdot(\vec{p}-\vec{k})\tau^a\eta\frac{1}{(p-k)^2-m_\pi^2},\label{tree}
\end{eqnarray}
where $\xi$ and $\eta$ are the spinors of the nucleon and antinucleon respectively.

At one-loop level, not all diagrams give contributions at the order
$\mathcal{O}(p^1)$. At one-loop level one meets a loop integral of a
four momentum $k$. One can perform the $k^0$-integration first by
using a contour in the complex $k^0$-plan. The contribution of the
integration comes from poles insider the contour. The poles can
appear from denominators of nucleon propagators or
$\pi$-propagators. It is easy to show that only those contributions from the poles
of nucleon propagators are at the order $\mathcal{O}(p^1)$, while other contributions
are at higher orders. With this in mind, only those diagrams at one-loop
given in Fig.2 give contributions at the next-to-leading order.

\par
\begin{figure}[hbt]
\begin{center}
  \includegraphics[width=12cm]{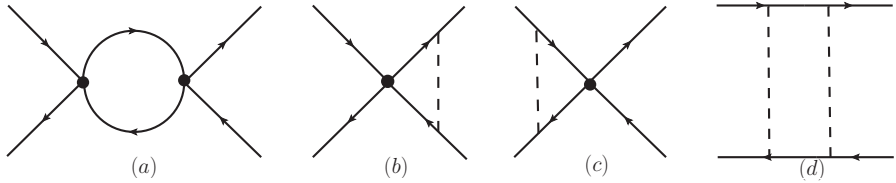}\\
  \caption{The diagrams which give the next-to-leading order corrections.}\label{fig2}
  \end{center}
\end{figure}
\par
It is rather straightforward to evaluate these loop contributions. In fact results of some loop integrals
exist in the literature. Hence, we give here the results without showing here detailed calculations.
It should be noted that all contributions in Fig.2 are finite in the dimensional regularization.
As discussed in the above, we will use the minimal subtraction scheme for renormalization. In this case,
the next-to-leading order contributions do not depend on the renormalization scale $\mu$, or the subtraction point.
The $\mu$-dependence of coupling constants starts at order of $p^2$.
The contributions from Fig.2a read:
\begin{equation}
\mathcal{T}\biggr\vert_{2a}=i \frac{mk}{4\pi}
\left [ \frac{c_0^2}{4}\xi^\dagger\eta\eta^\dagger\xi
+\frac{c_1^2}{4}\xi^\dagger\tau^a\eta\eta^\dagger\tau^a\xi
+\frac{d_0^2}{4}\xi^\dagger\sigma^i\eta\eta^\dagger\sigma^i\xi+
\frac{d_1^2}{4}\xi^\dagger\sigma^i\tau^a\eta\eta^\dagger\sigma^i\tau^a\xi \right ],
\end{equation}
we note that this contribution with  the minimal subtraction scheme can be equivalently calculated
by taking a cut cutting the nucleon loop in Fig.2a.
\par
The contributions from Fig.2b and Fig.2c read:
\begin{eqnarray}
\mathcal{T}\biggr\vert_{2b} & = & -i \frac{1}{4\pi}(\frac{g_A}{2f_\pi})^2\eta^\dagger\Gamma^{(c)}\xi
 \cdot \left[I_0(k)\xi^\dagger\sigma^i\tau^a\Gamma^{(c)}
\sigma^i\tau^a\eta+I_1(k)\xi^\dagger\vec{\sigma}\cdot\vec{k}\tau^a\Gamma^{(c)}\vec{\sigma}\cdot\vec{k}\tau^a\eta\right],
\nonumber\\
\mathcal{T}\biggr\vert_{2c}& =& -i \frac{1}{4\pi}(\frac{g_A}{2f_\pi})^2\xi^\dagger\Gamma^{(c)}\eta
\cdot \left[I_0(k)\eta^\dagger\sigma^i\tau^a
\Gamma^{(c)}\sigma^{i}\tau^a\xi+I_1(k)\eta^\dagger\vec{\sigma}\cdot\vec{p}\tau^a\Gamma^{(c)}\vec{\sigma}\cdot\vec{p}\tau^a\xi\right].
\end{eqnarray}
In the above $\Gamma^{(c)}$ is a matrix acting in the spin- and isospin space. The sum of the
direct product $\Gamma^{(c)}\otimes\Gamma^{(c)}$ is implied. The sum reads:
\begin{equation}
\Gamma^{(c)}\otimes\Gamma^{(c)}=\frac{c_0}{4}I\otimes
I+\frac{c_1}{4}\tau^a\otimes\tau^a+\frac{d_0}{4}\sigma^i\otimes\sigma^i+\frac{d_1}{4}\sigma^i\tau^a\otimes\sigma^i\tau^a.
\end{equation}
The functions $I_0$ and $I_1$ are from loop-integrals. They are:
\begin{eqnarray}
I_0(k)&=&\frac{mk}{4}\left[1+y-iy\sqrt{2y}+y(y+2)\left(-\frac{1}{2}\ln(1+\frac{2}{y})+i\tan^{-1}\sqrt{\frac{2}{y}}\right)\right],\nonumber\\
I_1(k)&=&\frac{m}{4k}\left[1-3y+3iy\sqrt{2y}-y(2+3y)\left(-\frac{1}{2}\ln(1+\frac{2}{y})+i\tan^{-1}\sqrt{\frac{2}{y}}\right)\right].
\nonumber\\
&& y=\frac{m_\pi^2}{2\vec{k}^2},\ \
k=\sqrt{\vec{k}^2}.
\end{eqnarray}
\par
The contributions from Fig.2d are the most difficult parts. However, the relevant results of loop integrals
can be found in \cite{Kaiser}. We have numerically checked these results and found an agreement.
The contributions can be written in the form:
\begin{eqnarray}
\mathcal{T}\biggr\vert_{2d}&=&m(\frac{g_A}{2f_\pi})^4 \biggr \{ F_S  \left ( 3 \xi^\dagger \xi \eta^\dagger \eta
 +2 \xi^\dagger \tau^a \xi \eta^\dagger \tau^a  \eta \right )
  -  \left ( 3 \xi^\dagger \sigma^ i \xi \eta^\dagger\sigma^j \eta
 +2 \xi^\dagger \sigma^i \tau^a \xi \eta^\dagger\sigma^j  \tau^a  \eta \right )
\nonumber\\
  && \cdot \left ( F_V \delta^{ij} + F_W (k-p)^i (k-p)^j + F_T (\vec{k}\times\vec{p})^i(\vec{k}\times\vec{p})^j \right )
\nonumber\\
     &&  + i F_P (\vec{k}\times\vec{p})^i \left (3 \xi^\dagger \sigma^i \xi \eta^\dagger \eta
      - 3 \xi^\dagger  \xi \eta^\dagger \sigma^i \eta
 +2 \xi^\dagger \sigma^i \tau^a \xi \eta^\dagger \tau^a  \eta
 -2 \xi^\dagger  \tau^a \xi \eta^\dagger \tau^a \sigma^i \eta\right ) \biggr\},
\end{eqnarray}
the functions $F_{S,V,T,W,P}$ are given as:
\begin{eqnarray}
F_S&=&-\frac{1}{16\pi
}\left\{4(2m_\pi^2+q^2)\Gamma_0(p)+2q^2\Gamma_1(p)-4ip-(2m_\pi^2+q^2)^2G_0(p,q)\right\},\nonumber\\
F_W&=&-\frac{1}{q^2}F_V=\frac{1}{4\pi}G_2(p,q),\nonumber\\
F_P&=&\frac{1}{8\pi}\left\{2\Gamma_0(p)+2\Gamma_1(p)-(2m_\pi^2+q^2)\left[G_0(p,q)+2G_1(p,q)\right]\right\},\nonumber\\
F_T&=&-\frac{1}{4\pi}\left\{G_0(p,q)+4G_1(p,q)+4G_3(p,q)\right\},\label{WV}
\end{eqnarray}
where the complex-valued functions are given as,
\begin{eqnarray}
\Gamma_0(p)&=&\frac{1}{2p}\left[\arctan\frac{2p}{m_\pi}+i\ln\frac{u}{m_\pi}\right],\
\ \ \ \ \ \ u=\sqrt{m_\pi^2+4p^2},\nonumber\\
\Gamma_1(p)&=&\frac{1}{2p^2}\left[m_\pi+ip-(m_\pi^2+2p^2)\Gamma_0(p)\right],\nonumber\\
G_0(p,q)&=&\frac{1}{q R}\left[\arcsin\frac{q m_\pi}{u
w}+i\ln\frac{pq+R}{u m_\pi}\right],\ \ \ R=\sqrt{m_\pi^4+p^2 w^2},\
\ \ w=\sqrt{4m_\pi^2+q^2},\nonumber\\
G_1(p,q)&=&\frac{\Gamma_0(p)-2A(q)-(m_\pi^2+2p^2)G_0(p,q)}{4p^2-q^2},\
\ \ \ A(q)=\frac{1}{2q}\arctan\frac{q}{2m_\pi},\nonumber\\
G_2(p,q)&=&p^2 G_0(p,q)+(m_\pi^2+2p^2)G_1(p,q)+A(q),\nonumber\\
G_3(p,q)&=&\frac{\frac{1}{2}\Gamma_1(p)-p^2
G_0(p,q)-2(m_\pi^2+2p^2)G_1(p,q)}{4p^2-q^2}.
\end{eqnarray}
where we define $q=|\vec{k}-\vec{p}|$ here.
The total results can be written as:
\begin{equation}
\mathcal{T}(\vec{p},\vec{k})=\mathcal{T}^{(0)}(\vec{p},\vec{k})+\mathcal{T}^{(1)}(\vec{p},\vec{k})+\mathcal{O}(p^2),\label{amp}
\end{equation}
where $\mathcal{T}^{(0)}$ is given in Eq.(\ref{tree}) and the order
$\mathcal{O}(p^1)$ result is:
\begin{equation}
\mathcal{T}^{(1)}=\mathcal{T}\biggr\vert_{2a}+\mathcal{T}\biggr\vert_{2b}
+\mathcal{T}\biggr\vert_{2c}+\mathcal{T}\biggr\vert_{2d}
\end{equation}
With the above results we have the complete result for
$N\bar N$ scattering near the threshold at the next-to-leading order.
We will use this result to fit the experimental data.
\par
\begin{figure}[hbt]
\begin{center}
  \includegraphics[width=10cm]{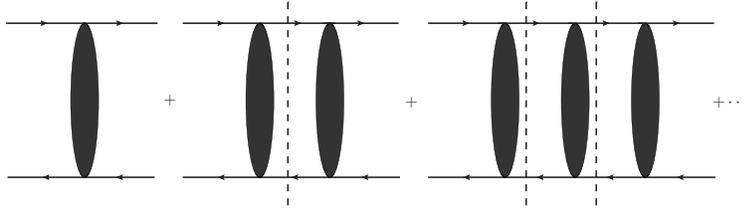}\\
  \caption{The narrow and long bubble represents the amplitude $\mathcal{A}_{[j\ell\ell's,I]}(E)$ at certain order
with the subtraction of contributions from physical cuts or cut diagrams.
The sum of diagrams can be performed.
} \label{Feynman-dg2}
  \end{center}
\end{figure}
\par
Our results for the scattering amplitude can be further improved by including some higher order corrections
without a concrete calculation.
We have noticed in \cite{CDM,CDM2} that some corrections from higher orders can be summed into a compact
form. The summation can be done for partial waves.
We define our partial waves from the $N \bar N$
scattering amplitude
$\mathcal{T}_I(\vec{p},\vec{k},s_1,s_2,s_1^\prime,s_2^\prime)$ with
the isospin $I=0,1$ as:
\begin{eqnarray}
\mathcal{T}_I(\vec{p},\vec{k},s_1,s_2,s_1^\prime,s_2^\prime)&=&\sum
Y^\ast_{\ell\ell_3}
(\vec{p}/p)Y_{\ell^\prime\ell_3^\prime}(\vec{k}/k)(1/2,s_1;1/2,s_2|s,s_3)(1/2,s_1^\prime;1/2,s_2^\prime|s^\prime,s_3^\prime)\nonumber\\
&&\cdot(\ell,\ell_3,s,s_3|j,j_3)(\ell^\prime,\ell_3^\prime,s^\prime,s_3^\prime|j^\prime,j_3^\prime)
\delta_{jj^\prime}\delta_{j_3j_3^\prime}\delta_{ss^\prime}\mathcal{T}_{[j\ell\ell^\prime
s,I]}(E).
\end{eqnarray}
In the above the repeated indexes are summed and $E$ is the total
kinetic energy $E=p^2/m$ of the system. The summation can be explained with Fig.3.
We will take $s$-waves as examples to illustrate this.
\par
Supposing we have calculated the scattering amplitude $\mathcal{T}_{[0000,0]}(E)$ at certain orders and we denote this
amplitude as $\mathcal{A}_{[0000,0]}(E)$, in which possible contributions from physical cuts or cut diagrams
are subtracted. At tree-level, $\mathcal{A}_{[0000,0]}(E)$ is just the tree-level amplitude.
At the next-to-leading order $\mathcal{A}_{[0000,0]}(E)$ is determined by diagrams from Fig.1 and those from Fig.2
after subtracting
the contributions of cut diagrams of Fig.2.
In Fig.3 we denote this amplitude with the narrow long bubble. Now using this amplitude
$\mathcal{A}_{[0000,0]}(E)$
one can generate those diagrams as ladder diagrams at higher orders as shown in Fig.3.
Physically the interpretation of Fig.3
is the following: The $N\bar N$ undergoes a multiple scattering process
$N\bar N \to N\bar N \to \cdots \to N\bar N$. Each scattering is due to the amplitude
$\mathcal{A}_{[0000,0]}(E)$
Each pair of $N\bar N$ is on-shell in Fig.3.
We will call amplitudes for such a multiple scattering process as rescattering amplitudes. It is easy to shown
that the sum of these diagrams is the sum of a geometric series. One then finds the sum for the amplitude
as:
\begin{equation}
{\mathcal T}_{[0000,0]}(E) = {\mathcal A}_{[0000,0]}(E)
\left [ 1 -i\frac{m^2\beta}{(4\pi)^2} {\mathcal A}_{[0000,0]}(E) \right]^{-1} + \cdots.
\end{equation}
This can be generalized to other partial waves. We will improve our result by adding the rescattering effects.
The improvement is done only for $S$-waves with the assumption that
the amplitudes with small $\ell$ are dominant because of the finite interaction range.
A small complication is with the case of $s=1$ is that there is a mixing of $S$-waves and $D$-waves.
In this case, the summation takes a form of $2\times 2$ matrix. Details can be found in \cite{CDM2}.

\begin{figure}[hbt]
\begin{center}
  \includegraphics[width=10cm]{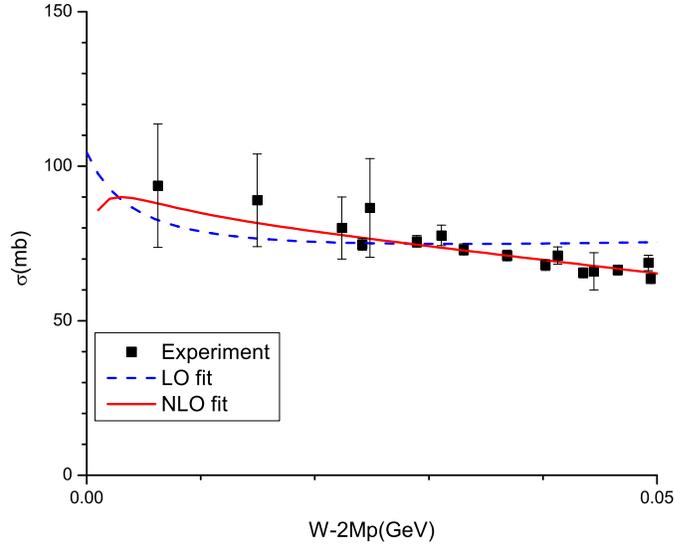}\\
  \caption{The fitting results for the cross-section of $p\bar p$ elastic scattering. The dashed line is the leading order fit.
  The solid line
  is the next-to-leading order fit.}\label{Graph1}
  \end{center}
\end{figure}

\par
We have performed numerical fits with our results for experimental data. The data are taken from \cite{PDG}.
In \cite{CDM2} we have studied the $p\bar p$ scattering and the enhancement
in $J/\psi \to \gamma p\bar p$
and $e^+ e^- \to p\bar p$ together with a combined fit.
For the enhancement, certain assumptions
have been made in \cite{CDM2} to relate the observed enhancement to $N\bar N$ scattering near the
threshold. In fact, the study of the enhancement can also be done with the approach
of effective theories in a consistent way combined with the effective theory here for $N\bar N$ system.
We leave this to future works.  Here we only focus on $p\bar p$ elastic scattering
in order to see how good our effective theory works.
In Fig.4.  we give our fitting result with the amplitude at leading and next-to-leading order, respectively.
With our results we are able to describe the experimental data below $E \sim 50$MeV, corresponding to $p\sim 200$MeV.
The fitting quality with the leading and the next-to-leading result is given by $\chi/d.o.f. \approx 2.6$
and $\chi/d.o.f. \approx 1.7$, respectively.
At tree-level we can not determine the coupling $c_{0,1}$ or $d_{0,1}$ separately, because
they appear as the combination as $c_0+c_1$ or $d_0+d_1$ in the amplitude. With the NLO results we can determine
these coupling constants in the unit of ${\rm GeV}^{-2}$ as:
\begin{eqnarray}
&& c_0 = (180 \pm 4 )+i(20 \pm 3), \ \ \  c_1=-211+i87,\ \ \
\nonumber\\
&& d_0=-(60 \pm 2)+i(61 \pm 4),
\ \ \  d_1=(37\pm 3)+i(56 \pm 4).
\end{eqnarray}
where the fitting errors are given in the brackets. For $c_1$ the errors are very small in comparison
with the numbers. We hence do not give the errors.
From Fig.4. one can see that our NLO amplitude can describe experimental data well.
With the determined couplings we can also determine the scattering lengthes of $S$-waves:
\begin{eqnarray}
&& a_{[^1S_0,0]} =-(5.88 \pm 0.22)-i(1.12 \pm 0.17), \ \ a_{[^1S_0,1]} =0.67-i0.09, \ \
\nonumber\\
&& a_{[^3S_1,0]} =4.24-i0.06, \ \ a_{[^3S_1,1]} =-(0.25\pm 0.06) -i(1.08\pm 0.08),
\end{eqnarray}
in unit of ${\rm fm}$. In the above the last integer in $[\cdots ]$ indicates the isospin $I=0$ or $I=1$.
\par
We have also performed the numerical fits by adding the rescattering effects into the leading- and next-to-leading
amplitude as discussed in the above. The fitting results are represented in Fig.5. From Fig.5. one can realize
that with the rescattering effects the experimental data can also be well described, indicated
by $\chi/d.o.f. \approx 1.7$.  From the fit
with next-to-leading amplitude and rescattering effects we have for the coupling constants in unit of ${\rm GeV}^{-2}$
as:
\begin{eqnarray}
&& c_0 = (70 \pm 37)+i(0\pm 190), \ \ \  c_1=(132 \pm 92) +i(0\pm 248),
\nonumber\\
&& d_0=-(25\pm 2)+i(0\pm 14) , \ \ \  d_1=-(38\pm 8)+i(99\pm 80).
\end{eqnarray}
The corresponding scattering lengthes are
\begin{eqnarray}
&& a_{[^1S_0,0]} =(0.3 \pm 2)-i(0 \pm 10), \ \ a_{[^1S_0,1]} =(0.34 \pm 0.09)-i(0.00 \pm 0.24), \ \
\nonumber\\
&& a_{[^3S_1,0]} =4.21-i(0\pm 0.01), \ \ a_{[^3S_1,1]} =(1.20\pm 0.15)-i(1.91\pm 1.54).
\end{eqnarray}
The determined scattering lengthes are qualitatively in agreement with results
from  some models\cite{Tbes3} and from LEAR experiment\cite{Tbes2}.
They are smaller or much smaller than those of an $NN$ system. This fact supports
the argument for using the minimal subtraction scheme and hence the standard power counting
for our effective theory.
\begin{figure}[hbt]
\begin{center}
  \includegraphics[width=10cm]{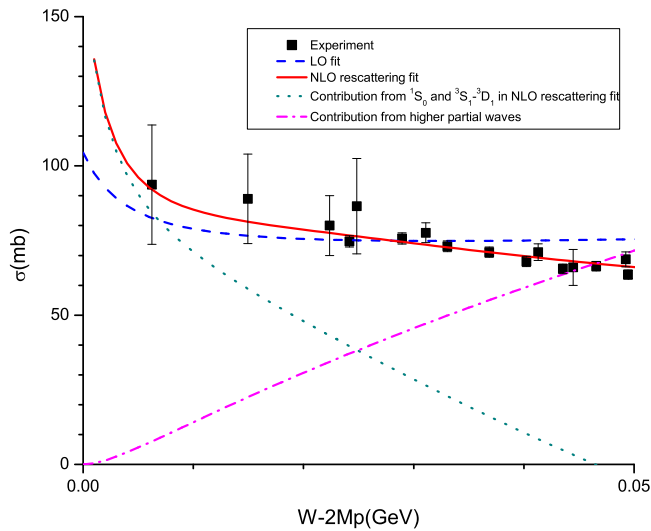}\\
  \caption{The fitting result for the cross-section of $p\bar p$ elastic scattering.
  The dashed line is the leading order fit. The solid line is the next to leading order fit while
  including
  the rescattering effect in $^1S_0$ and $^3S_1-{^3}D_1$ channel. The dotted line is the contribution from
  $^1S_0$ and $^3S_1-{^3D_1}$ with the rescattering amplitude. The dashed-dotted line is the contribution
  from all the other partial waves up to the next-to-leading order.}\label{Graph2}
  \end{center}
\end{figure}

\par
Although the data can be well described from Fig.5, the significant changes in some coupling constants
indicate that corrections from higher orders can be significant. In Fig.5 we also give results for the contributions
from those partial waves with $\ell\neq 0$ where the contribution from the partial wave $^3 D_1$ is subtracted.
These contributions essentially come from $\pi$-exchanges. From Fig.5. we can see that
these contributions will become dominant when the total energy $E$ is approaching to the value around
$50$MeV, corresponding to $p\sim 200$MeV. This indicates that the expansion in $p$ of the effective theory
converges slowly or applicable region of the effective theory is small. This has also been found
in the effective theory of an $NN$ system. The study in \cite{Fleming} shows that the contributions
from $\pi$-exchanges become dominant and much larger than predictions at leading orders  around $p\sim 100$MeV
in some spin triplet channels.
The reason for this is the true expansion parameter for $\pi$-exchanges
is at order of $(g_A^2 m_\pi m)/(8\pi f^2_\pi)$. Because the nucleon mass $m$ is large in comparison
with $f_\pi$, the expansion parameter is not small. In our case in which
contributions from $\pi$-exchanges become dominant at $p\sim 100$MeV,
the situation seems better than that for $NN$ systems, because the coupling constants of the local interactions,
which give the important contributions at leading order, are roughly twice larger than the corresponding
coupling constants in the effective theory of an $NN$-system.
\par
It is interesting to compare the approach of the effective theory with the approach of the
the effective range expansion for describing $N\bar N$ scattering. In the effective range expansion, e.g.,
in \cite{EFR}, the scattering amplitude, or more exactly, the phase-shift is expanded
in $p$ and only first two terms in the expansion  are kept.
The first term is the scattering length and the second term is determined
by the interaction range. These terms can be complex. If we neglect
interactions with $\pi$ in our effective theory, our effective theory will only
consist of local operators. The perturbative expansion of coupling constants of the
local operators can be summed into a compact form.
In the case, our approach will be identical to
that of the effective range expansion. However, the existence of interactions
with $\pi$ makes two approaches different. We notice that interactions with $\pi$
can not be represented with local operators. To use local operators for the interactions,
one in essence makes an expansion in $p/m_\pi$ for the contribution from Fig.1b.
But this expansion converges very slowly with the convergence range about several MeV, reflecting
the fact that the scattering amplitude through $\pi$-exchanges has a cut very near- and below
the threshold. In our approach we do not make such an expansion in $p/m_\pi$.

\par
To summarize: We have studied an effective theory for an $N\bar N$ system near the threshold.
The effective theory is constructed by an expansion of momenta near the threshold and including
interactions with $\pi$-mesons fixed with chiral symmetry.
The power counting of the effective theory has been given.
We have obtained the next-to-leading
order results for $N\bar N$ scattering and compared with experimental data. The data can be well fitted
with our predictions. But, there can be the problem that the expansion of our effective theory
does not converge quickly. This may need to be studied further. If there are
more experimental data near the threshold, our effective theory can be tested more accurately.
We notice that one can also predict the spin-dependent $N\bar N$ scattering by using our effective theory
and compare with results from the planned experiment\cite{PAX}. This will provide another
interesting test of our effective theory.

\par\vskip20pt

\vskip 5mm
\par\noindent
{\bf\large Acknowledgments}
\par
We would like to thank Prof. H.Q. Zheng and Dr. X.G. Wang for
interesting discussions. This work is supported by National Nature
Science Foundation of P.R. China(No. 11021092).
\par

\end{document}